%% file: ipdps/main.tex
\DocumentMetadata{}

\documentclass[10pt,conference]{IEEEtran}

\usepackage{cite}
\usepackage{amsmath,amssymb,amsfonts}
\RequirePackage{algorithm}
\RequirePackage{algorithmic}
\usepackage{graphicx}
\usepackage{textcomp}
\usepackage{xcolor}
\usepackage[hyphens]{url}
\usepackage{fancyhdr}
\usepackage{hyperref}
\usepackage{comment}
\usepackage{xspace}
\usepackage{multirow}
\usepackage{booktabs}
\usepackage{makecell}
\usepackage{capt-of}
\usepackage{multicol}
\usepackage[normalem]{ulem}
\usepackage{listings}
\usepackage{graphicx}

\def\BibTeX{{\rm B\kern-.05em{\sc i\kern-.025em b}\kern-.08em
    T\kern-.1667em\lower.7ex\hbox{E}\kern-.125emX}}

\newcommand{\name}{WGLog\xspace}

\title{Terascale Query Processing in the Browser: \emph{Rethinking GPU acceleration}}

\author{
\IEEEauthorblockN{
Jiaxin Lu\IEEEauthorrefmark{1},
Landon Dyken\IEEEauthorrefmark{1},
Yihao Sun\IEEEauthorrefmark{2},
Kristopher Micinski\IEEEauthorrefmark{3},
Thomas Gilray\IEEEauthorrefmark{4},
and Sidharth Kumar\IEEEauthorrefmark{1}
}
\IEEEauthorblockA{
\IEEEauthorrefmark{1}University of Illinois Chicago, Chicago, IL, USA, 
\{jlu73, ldyke, sidharth\}@uic.edu
}
\IEEEauthorblockA{
\IEEEauthorrefmark{2}Utah State University, Logan, UT, USA,
yihao.sun@usu.edu
}
\IEEEauthorblockA{
\IEEEauthorrefmark{3}Syracuse University, Syracuse, NY, USA,
kkmicins@syr.edu
}
\IEEEauthorblockA{
\IEEEauthorrefmark{4}Washington State University, Pullman, WA, USA,
thomas.gilray@wsu.edu
}
}

\begin{document}
\maketitle

\begin{abstract}
Recursive query computation, central to graph algorithms and relational databases, demands GPU acceleration due to inherent computational intensity. While substantial prior work addresses GPU implementations of recursive queries, requiring fixed-point evaluation, existing systems are restricted to native execution environments.
We introduce \name, the first web-browser-native GPU engine for compute-bound recursive database queries. \name is built entirely on WebGPU compute shaders, a cross-platform API that enables GPU acceleration in web browsers. \name leverages two key technical innovations. First, we replace hash-table-based joins with atomic-free sorted-array joins, eliminating the serialization bottleneck that hash tables suffer on skewed graphs. Second, we develop an asynchronous execution pipeline using WebGPU's indirect dispatch capability, which eliminates GPU-host synchronizations that would otherwise dominate per-iteration overhead.
On representative workloads, \name delivers 1.48–4.68× speedup over native GPU systems and orders of magnitude improvement over CPU and webAssembly implementations.

\end{abstract}

\input{ipdps/text/intro}
\input{ipdps/text/background}
\input{ipdps/text/motivation}
\input{ipdps/text/method1}
\input{ipdps/text/method2}
\input{ipdps/text/method3}

\input{ipdps/text/evaluation}
\input{ipdps/text/portability}

\input{ipdps/text/related_work}
\input{ipdps/text/conclusion}
\bibliographystyle{ipdps/IEEEtranS}
\bibliography{ref}

\end{document}

%% file: ipdps/text/intro.tex

\section{Introduction}
\label{sec:introduction}

Relational query processing with SQL excels at single-pass analytical workloads but struggles with recursion. SQL was designed for flat, acyclic computations, making recursive queries such as graph closure and transitive reachability cumbersome to express and inefficient to evaluate. Datalog~\cite{green2013datalog, huang2011datalog}, by contrast, is a simple but expressive declarative language purpose-built for recursive queries. It powers state-of-the-art systems across program analysis~\cite{bravenboer2009,souffle_cc2016}, network and graph analytics~\cite{seo2013,fan2021}, and knowledge-graph reasoning~\cite{urbani2016}.
These applications rely on recursive queries, a common computational pattern in which new facts are iteratively derived from existing data until saturating at a \emph{fixed point}, a state where further rule application produces no new tuples.

At scale, recursive queries become bandwidth-bound as iterations repeatedly stream through millions of tuples~\cite{kumar2019distributed}.
Mainstream engines such as Souffl\'e~\cite{souffle_cc2016} and DDlog~\cite{ddlog2019} run on CPUs and are limited by CPU memory bandwidth when closures reach tens of millions of tuples (80 million in the largest graph we evaluate~\cite{shovon2023,gpulog}).
GPUs are a natural alternative because their high-bandwidth memory and massive parallelism match this access pattern. 
Recent CUDA systems confirm this potential. State-of-the-art datalog engines, mnmgJOIN~\cite{shovon2023} and GPUlog~\cite{gpulog} accelerate TC and SG fixpoints, achieving order-of-magnitude speedups over Souffl\'e on large graphs.

Deploying portable implementations of Datalog outside CUDA environments remains an open problem. Both mnmgJOIN and GPUlog require NVIDIA-specific toolchains and produce native binaries, restricting deployment to systems with a CUDA driver installed. This limitation is increasingly costly as analytics shift toward on-device processing for privacy and compliance. Medical, financial, and personal-data workloads cannot upload to remote servers and must process locally. GPU computing has evolved dramatically. Once confined to specialized systems, GPUs are now ubiquitous~\cite{kirk2007nvidia}. Nearly all modern laptops contain powerful GPUs. Yet no infrastructure exists to exploit these GPUs within the \emph{portable web} environment. The web browser remains disconnected from GPU acceleration despite serving billions of users daily. A GPU-accelerated database engine in the browser would close this gap, democratizing high-performance analytics and enabling users to process large datasets locally on their existing hardware without specialized toolchains or expertise.



We address this gap with \name{}, the first GPU-accelerated database engine designed for browser execution, built entirely on WebGPU compute shaders. WebGPU~\cite{webgpu} provides a portable GPU compute API that runs in browsers without driver installation. Instead of being a port of CUDA implementation \name{} is designed from scratch. A direct port of existing CUDA pipelines to WebGPU fails for two fundamental reasons. First, a naive port would inherit the hash-join serialization bottleneck at hub vertices (high-degree) that already constrains CUDA baselines on skewed graphs~\cite{shovon2023,gpulog}. Second, WebGPU's record-submit-fence command model makes per-stage host synchronization expensive when repeated across hundreds of fixpoint iterations, unlike CUDA's approach. To address both problems, \name{} redesigns the recursive evaluation pipeline entirely. Rather than using hash tables, it replaces the hash-based pipeline with a sorted-array pipeline that avoids contention at hub vertices. Additionally, it exploits the observation that Datalog's fixpoint semantics require only a termination signal, not per-stage intermediate sizes~\cite{abiteboul1995}. \name{} batches multiple iterations into a single WebGPU command submission using indirect dispatch, a mechanism that lets the GPU compute its own dispatch parameters so the host never needs to read back intermediate sizes. We validate \name{} on three representative compute-bound workloads: transitive closure, same-generation queries, and triangle counting.


This paper makes the following contributions:
\begin{itemize}
    \item A sorted-array pipeline that replaces hash join with binary-search join and ordered set operations, eliminating data-skew-induced serialization (\S\ref{sec:sorted_set}).
    \item A batched indirect-dispatch fixpoint loop that reduces GPU-host synchronization across iteration, reducing non-compute overhead to 2--4\% of end-to-end time, compared with 68--70\% in CUDA baselines (\S\ref{sec:engine}).
    \item An implementation and evaluation of \name{} on 12~TC and 4~SG datasets, demonstrating $2.38\times$ cumulative speedup over mnmgJOIN, $3.05\times$ over GPULog, and $25.32\times$ over Souffl\'e on the same NVIDIA GeForce RTX 3060 Laptop GPU (\S\ref{sec:eval}).
\end{itemize}

To our knowledge, this is the first GPU-accelerated implementation of recursive database queries for the web browser. We compare against three WebAssembly ports of popular database engines (SQLite, DuckDB, and Ascent). As expected, \name{} significantly outperforms these baselines because WebAssembly does not leverage GPU acceleration. 

%% file: ipdps/text/background.tex

\section{Background}
\label{sec:background}

Datalog is a declarative logic language, where users specify \emph{what} facts to derive, not \emph{how} to derive them.
A program consists of rules of the form:
\[ H(\bar{x}) \leftarrow B_1(\bar{x}_1), \ldots, B_k(\bar{x}_k). \]
The \emph{head} $H$ is the fact to derive; the \emph{body} $B_i$ are conditions that must hold; shared variables express \emph{joins}.
Each rule body represents a relational-algebra expression where the conjunction of body atoms forms a join with equality conditions on shared variables.
Datalog extends relational algebra with a fixpoint operator that gives recursive rules their iterative semantics.
A rule whose body references the relation it defines is \emph{recursive}.
Datalog's semantics are based on \emph{fixpoint computation}, where rules are repeatedly applied to the database until no new fact is derived, reaching the \emph{fixpoint}~\cite{abiteboul1995}.
An important property of these semantics is that derived relations are \emph{sets}, so re-deriving an existing fact does not change the database.

\begin{figure}[t]
\centering
\vspace{-10pt}
\includegraphics[width=\columnwidth]{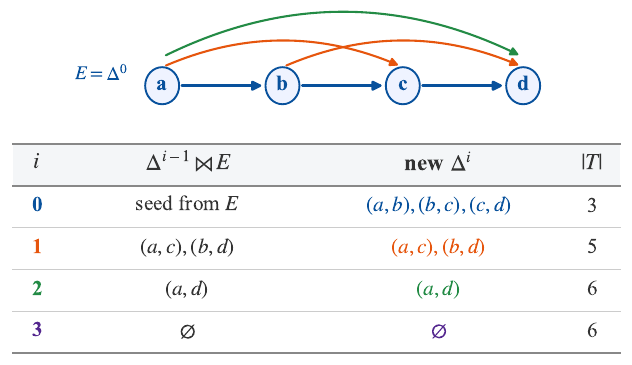}
\vspace{-15pt}
\caption{Transitive Closure on the running-example chain $a \to b \to c \to d$. \emph{Top:} the input graph contains only the blue edges $E$; orange paths (length 2) are derived in iteration 1 and the green path (length 3) in iteration 2. \emph{Bottom:} semi-na\"ive evaluation joins only the new delta $\Delta^{i-1}$ against $E$ at each iteration, producing $\Delta^i$ until it becomes empty.}
\label{fig:tc_example}
\vspace{-5pt}
\end{figure}

We illustrate fixpoint computation with Transitive Closure (TC), the running example used throughout the paper.
Given a directed graph with edge relation $E$, TC computes the reachability relation $T(x,y)$ using one base rule (left) and one recursive rule (right):
\[
  T(x, y) \leftarrow E(x, y).
  \qquad
  T(x, z) \leftarrow T(x, y), E(y, z).
\]
The base rule copies each edge into $T$; the recursive rule extends a known path by one hop.
Consider the four-vertex chain in Figure~\ref{fig:tc_example} (top).
The base rule seeds $T$ with the three input edges $(a,b)$, $(b,c)$, $(c,d)$.
A na\"ive evaluator then re-applies the recursive rule to every pair in $T$ at every iteration. \emph{Iteration~1} derives two new pairs: $(a,c)$ from $T(a,b) \wedge E(b,c)$, and $(b,d)$ from $T(b,c) \wedge E(c,d)$. Both are added to $T$. \emph{Iteration~2} again applies the rule to every pair currently in $T$, including $(a,b)$ and $(b,c)$. Because $E$ has not changed between iterations, joining $T(a,b) \wedge E(b,c)$ a second time produces $(a,c)$ again, and $T(b,c) \wedge E(c,d)$ produces $(b,d)$ again---these are duplicates of facts already in $T$. Only one derivation is genuinely new: $(a,d)$, obtained by joining the freshly-added pair $T(a,c)$ with $E(c,d)$.
Because $T$ is a set, the re-derivations of $(a,c)$ and $(b,d)$ contribute nothing. The engine performs the join, finds the result already in $T$, and discards it. \emph{Iteration~3} produces only such discarded re-derivations, and $T$ has reached its fixpoint of six pairs.

The source of redundancy is structural. Once a pair $(x,y)$ enters $T$, every subsequent iteration re-applies the rule to $(x,y)$ and re-derives every output it can produce. On real graphs with millions of pairs in $T$ and hundreds of iterations, this re-work dominates the cost.

\paragraph{Semi-Na\"ive Evaluation}
\label{sec:seminaive}

Semi-na\"ive evaluation exploits a key observation to eliminate redundancy. Each pair needs to join with $E$ only once, immediately after it is first derived. No subsequent iterations gain new information from re-joining existing pairs, as later iterations would only re-derive outputs that are already in $T$.
The algorithm therefore maintains two disjoint relations. A \emph{full relation} holding facts that have already been used as join input in earlier iterations, and a small \emph{delta} $\Delta^i$ holding the facts newly derived in iteration $i$ (waiting to be used as input in the next iteration).
The algorithm proceeds as follows:
\begin{enumerate}
  \item The base rule seeds $\Delta^0$ with the input edges.
  \item Iteration $i$ joins $\Delta^{i-1}$ against $E$, producing candidate tuples.
  \item From those candidates, the engine removes duplicates and subtracts those already in the full relation (set difference).
  \item The surviving candidates become $\Delta^i$ and are merged into the full relation.
  \item Evaluation terminates when $\Delta^i$ is empty.
\end{enumerate}

Figure~\ref{fig:tc_example} (bottom) traces this process on the running-example chain.
The base rule seeds $\Delta^0$ with the three input edges $(a,b)$, $(b,c)$, $(c,d)$, giving $|T|=3$.
\emph{Iteration~1} joins $\Delta^0$ against $E$ to produce $(a,c)$ and $(b,d)$; both are new, so $\Delta^1 = \{(a,c),(b,d)\}$ and $|T|=5$.
\emph{Iteration~2} joins $\Delta^1$ against $E$: only $(a,c)$ extends through $E(c,d)$ to yield $(a,d)$, since $(b,d)$ has no outgoing edge from $d$; thus $\Delta^2 = \{(a,d)\}$ and $|T|=6$.
\emph{Iteration~3} finds no extension for $(a,d)$ (which also dead-ends at $d$), so $\Delta^3 = \emptyset$ and the evaluation terminates. Each iteration joins only the latest delta against $E$ rather than the whole of $T$, so the per-iteration work scales as $|\Delta^{i-1}| \cdot |E|$ rather than $|T|^2$.

\subsection{GPU Parallelism in Recursive Evaluation}
\label{sec:gpu_background}

Each iteration of the semi-naïve loop comprises five stages: join, sort, deduplicate, set difference, and merge. All five stages are data-parallel, making them well-suited to GPU acceleration. State-of-the-art GPU systems mnmgJOIN~\cite{shovon2023} and GPULog~\cite{gpulog}, both built on NVIDIA CUDA, implement each stage as a distinct kernel. The join phase is the most compute-intensive and has become the focus of optimization. Both mnmgJOIN and GPULog implement the join using hash tables, following the classical hash join approach~\cite{shovon2023}. The base relation is inserted into a GPU-resident hash table, and each delta tuple probes it to find matching edges. 

Between iterations, the host reads back the delta size to decide whether fixed-point has reached or not. This read-back is inexpensive under CUDA; a single \texttt{cudaMemcpy} of four bytes takes ${\sim}9\,\mu s$ in our measurements (\S\ref{sec:eval}). Thus, per-iteration host coordination is essentially free, and the bottleneck is instead the hash join itself, which serialises on hub vertices in skewed graphs (\S\ref{sec:motivation}).
WebGPU, the portable alternative we target, departs from CUDA precisely on this assumption.

\subsection{WebGPU Compute}
\label{sec:webgpu}

WebGPU~\cite{webgpu} provides a portable GPU compute API that runs in browsers and native runtimes, with shaders written in WGSL~\cite{wgsl}.
Implementations map WebGPU commands to platform backends such as Vulkan~\cite{vulkan}, Direct3D~12~\cite{d3d12}, and Metal~\cite{metal}.
For recursive query evaluation, four aspects of WebGPU's execution model are relevant to our work.

\textbf{Dispatch and workgroups.}
The basic unit of work is a compute dispatch, which launches a grid of \emph{workgroups} (analogous to CUDA blocks).
Each workgroup contains a fixed number of invocations  (analogous to CUDA threads) that share on-chip memory and synchronise through barriers.
Invocations within a workgroup are further grouped into \emph{subgroups} (analogous to CUDA warps) that execute in lockstep and exchange values through intrinsics such as ballot (which gathers one bit from each invocation into a single mask) and broadcast (which propagates one invocation's value to all others).

\textbf{Indirect dispatch.}
In a standard dispatch, the host specifies the grid dimensions.
In an \emph{indirect dispatch}, the GPU reads them from a buffer that a prior kernel has written.
This mechanism is critical when the output size of one pipeline stage determines the grid size of the next, which is exactly the situation in each semi-na\"ive iteration.

\textbf{Command model.}
WebGPU follows a record--submit--fence model, where the host records dispatches into a command buffer, submits the buffer, and cannot observe any result until the entire buffer completes.
Reading a value back requires two additional steps: copying it to a CPU-mappable \emph{staging buffer}, and issuing an asynchronous \texttt{mapAsync} call that yields the host thread until the GPU finishes.
The resulting round-trip costs ${\sim}2.5\,ms$ in our measurements (\S\ref{sec:eval}), roughly 285$\times$ more than the equivalent CUDA \texttt{cudaMemcpy}.

\textbf{Buffer aliasing.}
WebGPU forbids a single buffer from appearing in two binding slots of the same compute pass when one slot is writable.
Algorithms that update data in place under CUDA must instead alternate between two buffers, using a \emph{ping-pong} pattern.

These constraints mean that porting a CUDA-style recursive engine to WebGPU by replacing API calls one-for-one would inherit two distinct bottlenecks. The first is the hash-join bottleneck from the algorithmic design, and second being the read-back bottleneck from the command model.
Together, these set the cost floor that \name{}'s redesign must beat.

%% file: ipdps/text/motivation.tex


\section{Motivation}
\label{sec:challenge}
\label{sec:motivation}

Existing GPU Datalog engines~\cite{shovon2023,gpulog} execute semi-na\"ive evaluation as a host-orchestrated pipeline of joins, sorting, duplicate removal, set difference, and merge.
This design assumes that hash-table accesses are close to uniform, that the host can cheaply sequence dependent GPU work, and that buffers can be updated in place.
Recursive graph workloads and WebGPU violate all three assumptions.
Data skew turns hash-based joins into serialized probe and retry chains (\S\ref{sec:data_skew}); data-dependent stage sizes force repeated host--GPU synchronization in a host-driven loop (\S\ref{sec:iter_overhead}); and WebGPU's no-aliasing rule rejects in-place buffer updates that CUDA pipelines rely on (\S\ref{sec:binding_motivation}).
The fixpoint loop repeats these costs across many iterations, so the performance problem is structural rather than an isolated slow kernel.

\begin{figure}[t]
\centering
\includegraphics[width=\columnwidth]{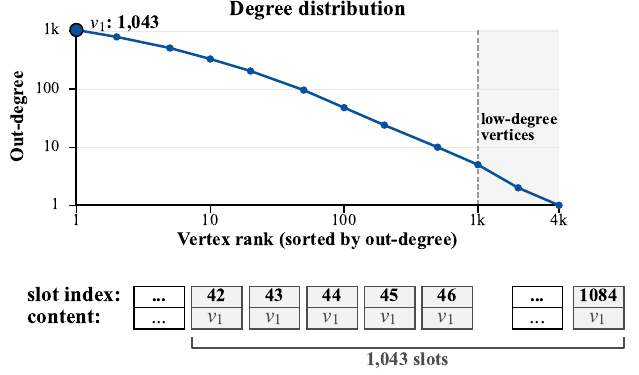}
\caption{Data skew in \texttt{ego-Facebook}. \emph{Top:} log-log plot of out-degree by vertex rank (sorted high-to-low); the highest-ranked vertex has $1{,}043$ outgoing edges. \emph{Bottom:} those $1{,}043$ edges all share the same hash key ($v_1$) and occupy $1{,}043$ consecutive slots in the hash table.}
\label{fig:data_skew}
\end{figure}
\subsection{Data Skew in Hash-Based Join}
\label{sec:data_skew}
\label{sec:hub_contention}

The first issue is an algorithmic cost within the join kernel itself.
Hash-based GPU joins are attractive because each delta tuple can probe the hash table independently.
The assumption behind this design is that join keys distribute evenly enough for linear probing to remain close to constant time.
Real graph inputs often violate that assumption: degree distributions are skewed, and a small number of vertices have out-degrees orders of magnitude larger than the median (Figure~\ref{fig:data_skew}).
In TC, all outgoing edges of a high-degree vertex share the same join key, so lookup and insertion requests for that key concentrate in the same hash-table region.

\begin{figure}[t]
\centering
%
%
%
%
%
\includegraphics[width=\columnwidth]{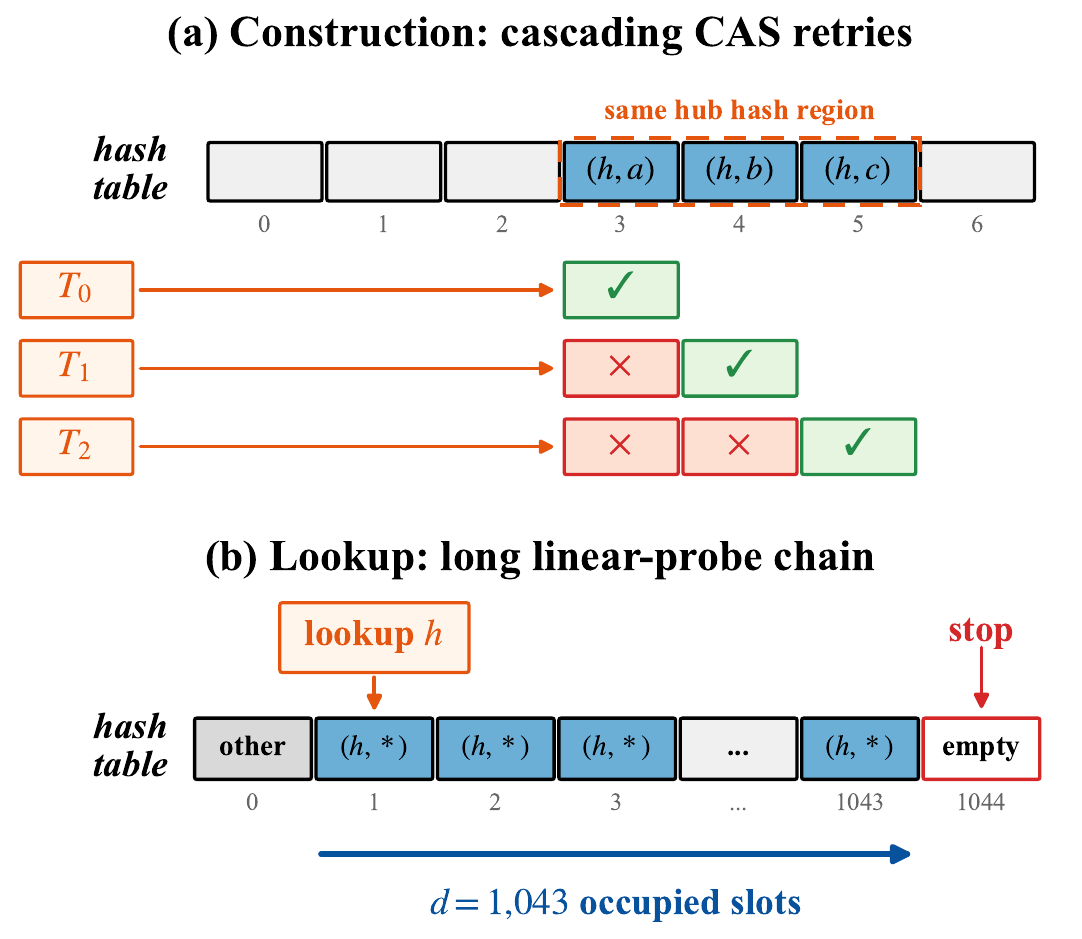}
\vspace{-0.8cm}
\caption{Two serialisation mechanisms at a hub vertex of out-degree $d$. \emph{Top (construction):} $d$ concurrent threads insert edges that share the same hash key; each thread fails CAS on every already-claimed slot and retries the next, so the $n$th thread does $n{-}1$ retries before succeeding, producing a cascade with $O(d^2)$ total CAS attempts. \emph{Bottom (lookup):} a probe for the hub key must walk all $d$ consecutive occupied slots one at a time, because each slot's load address depends on whether the previous slot matched the target key, blocking the GPU from issuing concurrent memory requests.}
\label{fig:hub_mechanism}
\vspace{-10pt}
\end{figure}

This concentration creates two serialization mechanisms (Figure~\ref{fig:hub_mechanism}), one during hash-table construction, and one during lookup.
During hash-table construction, threads inserting keys that map to the same neighborhood race to claim overlapping slots with atomic compare-and-swap (CAS).
Each failed CAS advances a thread to the next slot, producing a retry chain whose length grows with the local cluster size (number of keys hashed to the same slot).
During lookup, even a read-only probe must walk the same cluster through a chain of dependent memory loads. Each slot access reveals whether the target key has been found or the probe must continue to the next slot, so the address of each load depends on the result of the previous one.
The GPU can therefore issue only one outstanding memory request per thread in the skewed region, rather than the many concurrent requests it would issue under uniform keys.

On \texttt{ego-Facebook}~\cite{snap}, one vertex has out-degree $1{,}043$, making the worst-case probe path scale with the largest degree rather than the average degree.
GPULog rebuilds its hash index every iteration and pays both construction and lookup costs repeatedly.
mnmgJOIN builds the hash table once and avoids repeated construction, but every probe of the hub vertex still walks the long lookup chain in every iteration.
\name{} addresses this issue with a hash-free sorted-array pipeline (addressed in \S\ref{sec:sorted_set} and evaluated in \S\ref{sec:eval_ablation_sorted}).

\subsection{Stage-Wise Synchronization in Host-Driven Fixpoints}
\label{sec:iter_overhead}

\begin{figure}[t]
\centering
\includegraphics[width=\columnwidth]{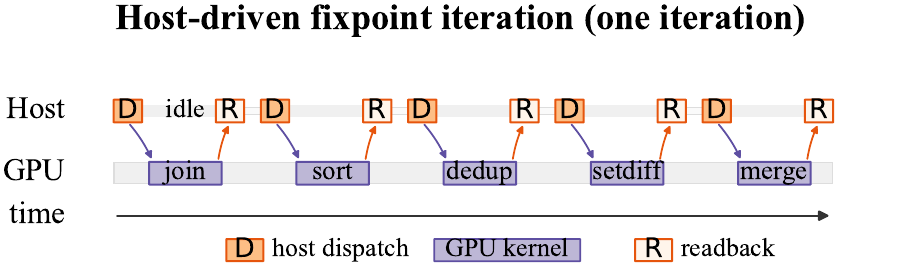}
%
%
\caption{Host-driven fixpoint pipeline of existing GPU Datalog engines: ${\sim}10$ host--GPU synchronisation points per iteration in both mnmgJOIN and GPULog.}
\label{fig:pipeline_sync}
\vspace{-10pt}
\end{figure}

The second issue is a system-level cost between kernels rather than within them.
Each semi-na\"ive iteration contains several dependent phases, and each phase produces an output size needed by the next phase (Figure~\ref{fig:pipeline_sync}).
In a host-driven implementation, the host must observe these device-produced sizes before it can allocate the next buffer, call a library primitive, or launch the next grid with the correct dimensions.
These observations are synchronization points, where the host waits for the GPU, reads a size value, and then issues the next dependent operation.

The cost is significant because recursive queries repeat the phase sequence many times.
A source-level audit gives approximately 13 synchronization points per iteration for mnmgJOIN and 10--14 for GPULog.
On three graphs requiring $188$--$426$ iterations to converge, mnmgJOIN's GPU kernels account for only $30$--$32\%$ of end-to-end time.
The remaining 68--70\% is non-compute overhead: kernel launch latency, host--GPU synchronization, and implicit buffer allocations within Thrust~\cite{thrust} and CUB~\cite{cub} library calls.
GPULog's non-compute fraction is lower, 21--34\%, because its per-iteration GPU work is heavier, but the absolute overhead remains substantial.
On \texttt{SF.cedge}~\cite{suitesparse}, which takes $287$ iterations and produces an $80$M-tuple closure, thousands of host-mediated stage transitions occur in one query.

WebGPU makes this issue central because its command model is record--submit--fence rather than a stream of cheap host-visible kernel launches.
However, Datalog's declarative semantics do not require the host to observe every intermediate stage size.
Within one iteration, and even across several iterations, the only value the host ultimately needs for control flow is whether the delta becomes empty.
\name{} addresses this issue with a batched indirect-dispatch engine that keeps intermediate sizes in GPU buffers and chains dependent stages without per-stage host readback (addressed in \S\ref{sec:engine} and evaluated in \S\ref{sec:eval_ablation_batch}).

\subsection{WebGPU Binding Rules Constrain In-Place Updates}
\label{sec:binding_motivation}

The third issue is WebGPU's buffer binding model.
CUDA implementations can often read and write the same global-memory allocation inside one kernel or across tightly coupled library calls.
WebGPU is stricter, where the same buffer range cannot serve as both the input binding and the output binding of a single compute pass~\cite{webgpu}.
For a recursive Datalog pipeline, this rule affects every stage that transforms a relation in place, including filtering, radix-sort passes, and merge.

A direct copy-back design would preserve the single-buffer abstraction, where each stage writes a temporary output and then copies it back into the input buffer expected by the next stage.
That design is simple but bandwidth-expensive, and recursive queries repeat the cost across many iterations.
The alternative is a ping-pong design that alternates between two physical buffers.
Ping-pong avoids the extra copy but consumes more memory and forces every stage to track which side currently holds the valid relation.
\name{} adopts ping-pong split arrays, parity-aware radix routing, and adaptive radix width as engine-level optimizations that make the sorted pipeline practical under WebGPU's binding rules (addressed in \S\ref{sec:engine} and evaluated in \S\ref{sec:eval_ablation_eng}).

\subsection{Implication for Design}
\label{sec:design_implications}

Taken together, the above three issues make clear that a browser-native Datalog engine cannot simply port the CUDA hash-based recipe.
The join path should avoid mutable hash-table state under skewed keys, the fixpoint loop should avoid host observation between dependent stages, and buffer management should respect WebGPU's no-aliasing rule without adding copy-back traffic.
\name{} therefore makes three design choices.
First, it represents recursive relations as sorted arrays and implements the recursive join through binary-search range lookup and ordered set operations (\S\ref{sec:sorted_set}).
Second, it records batches of dependent WebGPU dispatches using indirect dispatch arguments stored in GPU buffers, so the host checks termination only at batch boundaries (\S\ref{sec:engine}).
Third, it uses ping-pong split arrays and parity-aware sort routing to avoid illegal read-write aliasing without a copy-back stage (\S\ref{sec:engine}).
\S\ref{sec:eval_ablation_sorted} and \S\ref{sec:eval_ablation_batch} evaluate the algorithmic choices in isolation, while \S\ref{sec:eval_ablation_eng} measures the buffer-management tradeoff directly.
    

%% file: ipdps/text/method1.tex

\section{A Sorted-Array Pipeline within \name{} for Data-Skew-Resistant Joins}
\label{sec:sorted_set}

As introduced in \S\ref{sec:background}, Semi-na\"ive evaluation iterates over a five-stage loop, which includes, join, sort, deduplicate, set difference and merge, until the delta becomes empty (fixed point reached).
This section presents three sorted-array primitives that together implement the algorithmic stages without hash-table mutation, (a) a binary-search join for the join stage (\S\ref{sec:binsearch_join}), (b) a fused kernel combining deduplication and set difference (\S\ref{sec:fused_kernel}), and (c) a disjoint merge that incorporates the new delta into $\mathit{Full}$ (\S\ref{sec:disjoint_merge}).


\subsection{Binary-Search Join}
\label{sec:binsearch_join}
\S\ref{sec:data_skew} showed that skewed graph degrees turn hash-based joins into CAS retry chains during construction and long dependent probe chains during lookups.
\name{} gets rid of this bottleneck entirely by replacing hash-table lookup with sorted-array range queries.
The key invariant is simple. The base, delta, and full relations are lexicographically sorted at every stage boundary.
Sorting groups all tuples with the same join key into one contiguous range, so a lookup becomes a binary search followed by a sequential scan over the exact matches rather than a walk through a hash-table collision cluster.

Looking at the TC query, the recursive join pairs each newly discovered tuple $(a,b)$ in the delta with every base edge $(b,c)$ in $E$, producing candidates $(a,c)$. 
\name{} sorts $E$ once at dataset load and never modifies it, so all edges whose source is $b$ occupy one contiguous range.
For each delta tuple, the join binary-searches for the start of that range and scans forward through the matches.

\begin{figure}[t]
\centering
\includegraphics[width=\columnwidth]{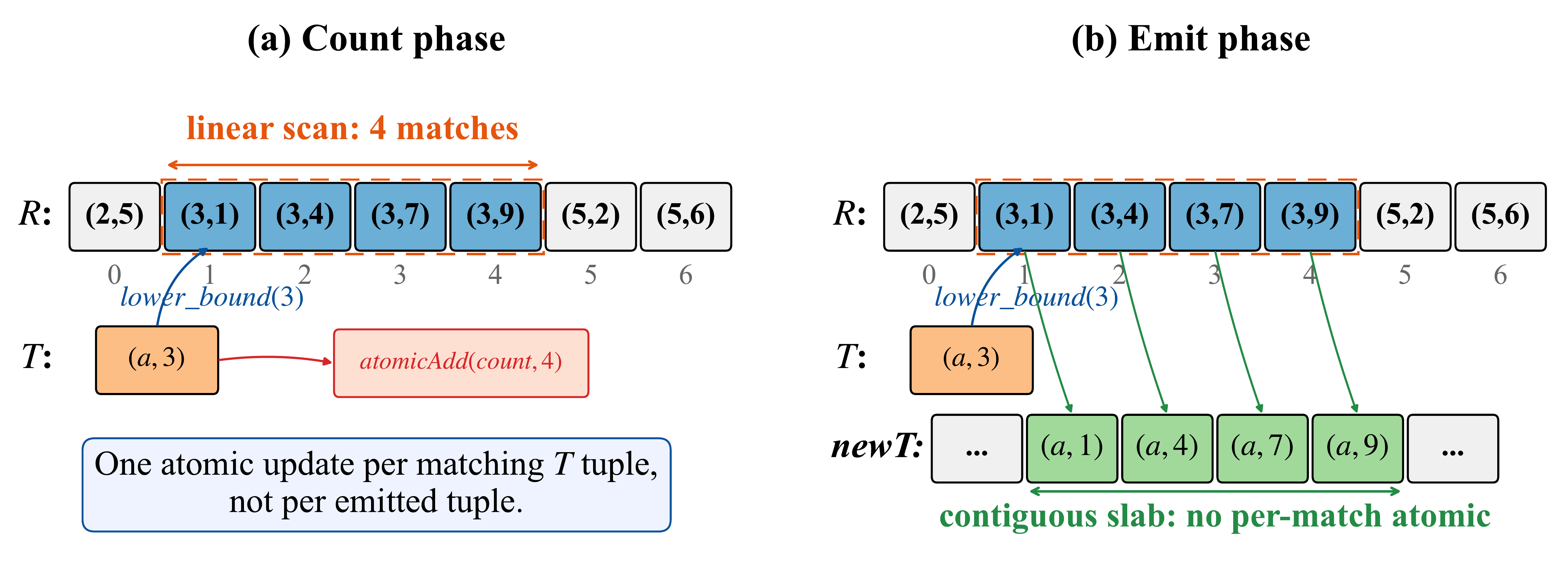}
\vspace{-15pt}
\caption{Binary-search join for delta tuple $(a,3)$: count phase locates the matching range in $E$; emit phase reserves one contiguous output slab and writes the matches without per-match atomics.}
\label{fig:binsearch_join}
\vspace{-10pt}
\end{figure}

\textbf{Two-phase algorithm.}
A delta tuple matching a vertex of out-degree $d$ emits $d$ candidates, so the total output size is unknown until the join runs.
\name{} therefore splits the join into a count phase and an emit phase (Figure~\ref{fig:binsearch_join}).
In the count phase, each thread handles one delta tuple, finds its matching range in $E$, and contributes the range length to a global candidate count.
In the emit phase, the thread repeats the same range lookup, reserves one contiguous output slab, and writes all candidates into that slab.
The repeated search is the tradeoff for avoiding per-output coordination, where the number of coarse reservations depends on the number of delta tuples, not on the number of emitted matches.

\textbf{Binary search avoids the hub-vertex bottleneck.}
A skewed vertex with out-degree $d$ creates a hash probe region whose cost grows with $d$ (\S\ref{sec:data_skew}); on \texttt{ego-Facebook}~\cite{snap}, the largest such vertex has $d = 1{,}043$.
By contrast, binary search reaches the matching range in at most $\lceil \log_2 |E| \rceil \le 18$ steps on our largest dataset ($|E| = 223{,}001$), after which the scan touches the $d$ matching edges contiguously.
The tradeoff is that low-degree vertices still pay the logarithmic search cost.
\name{} accepts the low-degree overhead because power-law graphs concentrate much of the join work at high-degree vertices; in our ablation, replacing hash lookup with binary search improves count-and-join time by $4.4\times$ on \texttt{ego-Facebook} (\S\ref{sec:eval_ablation_sorted}).

\subsection{Fused Deduplication and Set Difference}
\label{sec:fused_kernel}

After the binary-search join, the resulting candidates are sorted before becoming the next delta.
Sorting creates two useful properties: duplicate candidates become adjacent, and the full relation becomes searchable because it is also sorted by invariant.
The two filters required by semi-na\"ive evaluation therefore become local predicates over the candidate stream: keep a tuple only if it is the first copy among adjacent equal candidates and absent from the full relation.

\begin{figure}[t]
\centering
\vspace{-10pt}
\includegraphics[width=\columnwidth]{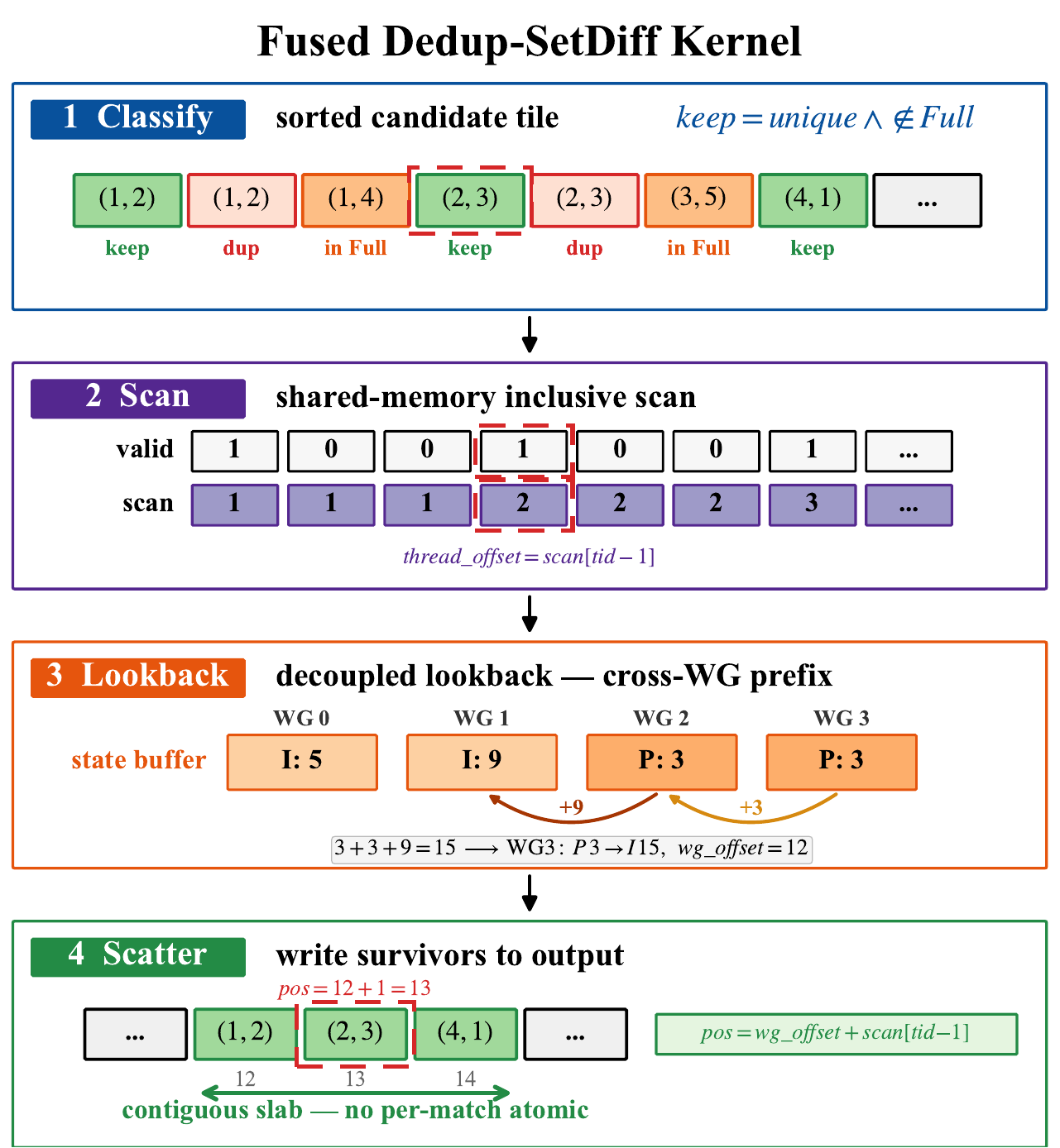}
\caption{The fused deduplication and set-difference kernel: classify, scan, decoupled lookback, scatter (\S\ref{sec:fused_kernel}).}
\label{fig:fused_kernel}
\vspace{-5pt}
\end{figure}

\textbf{Fusing the filters.}
A direct implementation materializes a deduplicated intermediate array and then reads it again for set difference.
\name{} avoids that round trip by fusing both filters into one tiled kernel (Figure~\ref{fig:fused_kernel}).
The first dispatch computes launch parameters from the candidate count; the second dispatch streams the sorted candidates once, evaluates both predicates, and emits only survivors.
Avoiding the intermediate array removes one global-memory round trip and reduces the separate-filter path from seven dispatches to two.

\textbf{Tiled execution.}
The kernel processes candidates in fixed-size tiles of $3{,}072$ tuples ($256$ threads $\times$ $12$ tuples per thread).
Within each tile, threads classify candidates, run a workgroup prefix scan to assign local write offsets, use decoupled lookback~\cite{merrill2016} to obtain the global tile offset, and scatter survivors in input order.
Because the input is sorted and each tile writes survivors in the same order, the output remains sorted for the next stage.
In our ablation, this fusion cuts cumulative end-to-end time by another $5.4\%$ (\S\ref{sec:eval_ablation_eng}).

\subsection{Disjoint Merge}
\label{sec:disjoint_merge}

After the fused kernel produces the new delta, that delta must be merged into the full relation while preserving the sorted-set invariant.
A standard Merge Path parallel merge~\cite{green2012gpu} combines two sorted arrays, but it does not remove duplicates: if the same tuple appears in both inputs, both copies are emitted.
\name{} avoids a post-merge deduplication pass because the fused kernel has already subtracted the full relation from the new delta.
The two inputs to the merge are therefore disjoint by construction.

The disjointness guarantee makes the merge simple.
The output size is exactly $|\Delta| + |\mathit{Full}|$, and each workgroup's output range is determined by its merge partition rather than by a data-dependent survivor count.
The merge therefore needs no per-tuple atomics and no cross-workgroup synchronization.

\textbf{Disjoint inputs eliminate the post-merge cleanup pass.}
Semi-na\"ive evaluation guarantees that $\Delta^i$ contains only tuples absent from the full relation before iteration~$i$.
Because the fused kernel enforces that guarantee, the merge inputs are disjoint, and the output remains a sorted set without a cleanup pass.
The WebGPU buffer-routing needed to implement this merge without read-write aliasing is part of the engine design in \S\ref{sec:engine}.

%% file: ipdps/text/method2.tex

\section{\name{}'s asynchronous Fixpoint Engine}
\label{sec:engine}

The sorted-array pipeline of \S\ref{sec:sorted_set} removes the data-skew bottleneck, but it does not by itself make the engine WebGPU-native.
\S\ref{sec:iter_overhead} showed that host-driven stage transitions create repeated synchronization, and \S\ref{sec:binding_motivation} showed that WebGPU's binding model prevents CUDA-style in-place updates.
\name{} therefore follows one engine rule: \emph{the host decides when the fixpoint terminates, but the GPU decides how each stage, iteration, and buffer transition proceeds}.

The engine implements this rule with three mechanisms.
\S\ref{sec:pipeline} describes the device-resident data flow of one iteration.
\S\ref{sec:batched_loop} shows how indirect dispatch removes host observation between dependent stages and batches multiple iterations into one submission.
\S\ref{sec:buffer_routing} explains the ping-pong and radix-routing scheme that satisfies WebGPU's no-aliasing rule without copy-back traffic.

\subsection{Per-Iteration Pipeline}
\label{sec:pipeline}

\begin{figure}[t]
\centering
%
%
\vspace{-5pt}
\includegraphics[width=0.75\columnwidth]{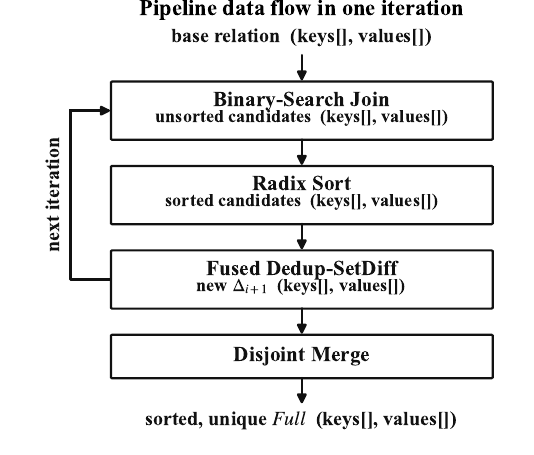}
\vspace{-0.2cm}
\caption{Data flow of one fixpoint iteration: split arrays of $32$-bit keys/values throughout, with no format conversion at stage boundaries. Deduplication and set difference (introduced as two separate stages are fused into a single kernel in our optimized implementation.}
\label{fig:pipeline}
\vspace{-8pt}
\end{figure}

The four stages of \S\ref{sec:sorted_set}, binary-search join, radix sort, fused deduplication and set difference, and disjoint merge, compose into one fixpoint iteration (Figure~\ref{fig:pipeline}).
The join produces unsorted candidate tuples, the radix sort restores lexicographic order, and the two set-operation stages then consume sorted arrays directly.
All stages operate on parallel 32-bit arrays of source and destination identifiers rather than packed 64-bit tuples, so no pack/unpack conversion is needed at stage boundaries. The important engine property is that every stage boundary is represented by buffers, not by host-visible values. Candidate counts, filtered-delta counts, merge sizes, and indirect-dispatch arguments all remain in device memory until a batch finishes.

\subsection{Batched Indirect-Dispatch Loop}
\label{sec:batched_loop}

\S\ref{sec:iter_overhead} attributed two-thirds of mnmgJOIN's end-to-end time to host--GPU synchronization: each stage's output size is unknown until the stage finishes, and the host needs that size before it can dispatch the next stage.
\name{} eliminates this dependency by routing every intermediate size through GPU-resident buffers; the host never reads any intermediate value during a batch.

\textbf{Indirect dispatch.}
Every kernel writes its output count to a $4$-byte device buffer.
A lightweight kernel reads that count and computes the dispatch parameters for the next stage, writing them into a second device buffer. The next stage then launches via \emph{indirect dispatch}, reading its launch configuration from that buffer rather than from the host. Because both the count and the dispatch parameters live on the device, no host--GPU synchronization occurs between consecutive stages within an iteration. The same pattern carries the new delta's size across iteration boundaries. 

\textbf{Batched recording.}
Because every dispatch within and across iterations reads its parameters from device memory, the host can record $K$ consecutive iterations into a single command buffer and submit them as one unit.
The GPU executes all $K$ iterations in sequence without returning to the host; only on completion does the host read back the per-iteration delta sizes to detect whether any reached zero (the fixpoint signal). On \texttt{SF.cedge}~\cite{suitesparse}, which converges in $287$ iterations, mnmgJOIN performs $287 \times 13 \approx 3{,}700$ host--GPU round trips per query.
With the transitive-closure default $K_{\max}=30$, a fixed-size batch schedule would reduce the loop to $\lceil 287/30 \rceil = 10$ termination checks before any adaptive shrink adjustments.
The command-model ablation in \S\ref{sec:eval_ablation_batch} isolates the resulting effect.

\textbf{Adaptive batch sizing.}
A larger $K$ amortizes the per-batch submission cost over more iterations, but risks \emph{overshoot}: if the fixpoint converges at iteration~$i$, the remaining $K - (i \bmod K)$ iterations dispatch on an empty delta and waste GPU time.
The waste dominates on short-fixpoint datasets where $K$ may exceed the total iteration count.
For transitive closure, \name{} starts from $K_{\max}=30$ and halves the next batch size whenever a completed batch shows strong delta decay (last-to-first size ratio below $0.1$).
The halving is monotonic, so near convergence the batch size shrinks to one or two iterations and limits the final empty work.
For same-generation queries, whose evaluated fixpoints run for only $54$--$77$ iterations, \name{} uses a fixed $K=10$ because the adaptive policy adds little benefit on these short loops.

\textbf{Double-buffered submission.}
The host hides recording latency by building batch $N{+}1$ while the GPU executes batch $N$; on batch $N$'s fence, the pre-recorded batch is submitted (if no termination is detected) or discarded.

\subsection{WebGPU Buffer Routing and Adaptive Radix Sort}
\label{sec:buffer_routing}
\label{sec:radix_sort}

\S\ref{sec:binding_motivation} explained that WebGPU rejects read-write aliasing within a compute pass.
The engine therefore treats every relation as a pair of physical buffers and flips roles after each stage that logically updates the relation.
This ping-pong scheme turns an illegal in-place update into legal read-from-A/write-to-B or read-from-B/write-to-A passes.
The cost is extra memory, but the benefit is avoiding a copy-back after every transformation.
The evaluation measures this tradeoff directly in \S\ref{sec:eval_ablation_eng}.

The fused kernel of \S\ref{sec:fused_kernel} requires its input to be lex-sorted on $(\textit{src}, \textit{dst})$, and the disjoint merge preserves this invariant on the full relation.
The join's output, however, is unsorted, so every iteration must sort its candidate array before the fused kernel can read it.
\name{} performs this sort with a GPU radix sort that processes eight bits per pass (histogram, prefix sum, scatter).
Because each radix pass reads one buffer pair and writes the other, radix sort is the stage most sensitive to ping-pong routing.

\textbf{Adaptive bit-width selection.}
A $32$-bit key requires four passes, a $24$-bit key three, a $16$-bit key two---and since each pass reads and writes the whole array, fewer passes translate directly into less memory traffic.
At dataset load, \name{} scans the maximum vertex identifier and picks the narrowest width that covers it: $16$ bits below $2^{16}$, $24$ bits below $2^{24}$, $32$ bits otherwise.
In the six-dataset sort ablation, four datasets use $16$-bit sorting and two use $24$-bit sorting; none requires the full $32$ bits.
Compared with a fixed $32$-bit radix, adaptive selection reduces sort time by $1.57\times$ on average, while the end-to-end gain is smaller because sort is not the dominant stage on every dataset (\S\ref{sec:eval_ablation_eng}).

\textbf{Parity-aware buffer routing.}
Lexicographic sorting first orders by the destination field (the secondary key) and then by the source field (the primary key), with each field requiring two, three, or four eight-bit radix passes depending on the selected width.
Each pass reads one split-array pair and writes another, so the implementation ping-pongs between an A-side and a B-side pair.
Because $24$-bit fields use an odd number of passes, the final output side would otherwise differ from the $16$- and $32$-bit cases.
\name{} routes the second field's input and output buffers according to this parity so the sorted candidates always end on the A-side pair, letting downstream stages read from fixed buffers.

%% file: ipdps/text/method3.tex
\begin{table*}[t]
\vspace{-15pt}
\centering
\scriptsize
\caption{TC end-to-end time (ms) on 12 graphs. $|\mathit{TC}|$ is the closure size in tuples; speedups below $1\times$ indicate \name{} is slower.}
\label{tab:tc_e2e}
\small
\resizebox{\textwidth}{!}{
\begin{tabular}{l r r r r r c}
\hline
Dataset & Iter. & $|\mathit{TC}|$ & \name{} & mnmgJOIN & GPULog & speedup vs.\ mnmgJOIN / GPULog \\
\hline
\texttt{OL.cedge}        &  64 &    146,120 &   68.538 &   54.83 &   94.78 & 0.80$\times$ / 1.38$\times$ \\
\texttt{cal.cedge}       & 195 &    501,755 &  155.366 &  332.48 &  318.50 & 2.14$\times$ / 2.05$\times$ \\
\texttt{TG.cedge}        &  58 &    481,121 &   55.551 &   83.88 &  103.32 & 1.51$\times$ / 1.86$\times$ \\
\texttt{p2p-Gnutella09}  &  20 & 21,402,960 &  239.319 &  313.51 & 1131.98 & 1.31$\times$ / 4.73$\times$ \\
\texttt{p2p-Gnutella04}  &  26 & 47,059,527 &  537.248 &  789.75 & 2654.20 & 1.47$\times$ / 4.94$\times$ \\
\texttt{cti}             &  53 &  6,859,653 &  127.389 &  252.23 &  401.28 & 1.98$\times$ / 3.15$\times$ \\
\texttt{fe\_sphere}      & 188 & 78,557,912 & 1965.439 & 4972.56 & 5778.39 & 2.53$\times$ / 2.94$\times$ \\
\texttt{ego-Facebook}    &  17 &  2,508,102 &  149.445 &  164.39 &  547.17 & 1.10$\times$ / 3.66$\times$ \\
\texttt{luxembourg\_osm} & 426 &  5,022,084 &  598.227 & 1208.42 & 1375.92 & 2.02$\times$ / 2.30$\times$ \\
\texttt{wing}            &  11 &    329,438 &   38.758 &   15.50 &   28.29 & 0.40$\times$ / 0.73$\times$ \\
\texttt{delaunay\_n16}   & 101 &  6,137,959 &  208.760 &  405.00 &  500.34 & 1.94$\times$ / 2.40$\times$ \\
\texttt{SF.cedge}        & 287 & 80,498,014 & 2467.678 & 7180.45 & 7211.61 & 2.91$\times$ / 2.92$\times$ \\
\hline
\end{tabular}}
\vspace{-5pt}
\end{table*}
\section{Generalization Across Datalog Queries}
\label{sec:query_coverage}

Datalog can express a wide range of queries beyond transitive closure, including graph reachability, ancestor relationships, social-network analyses, knowledge-graph reasoning, and triangle/clique enumeration~\cite{abiteboul1995,bravenboer2009,seo2013,fan2021,urbani2016}. The \name{} engine described in \S\ref{sec:sorted_set}--\ref{sec:engine} is not specific to TC; the sorted-array primitives and the batched indirect-dispatch loop apply to any query expressible as a fixpoint over join, set operations, and merge over sorted relations.
This section illustrates the generality with two additional queries: \emph{same-generation}, and \emph{triangle counting}.

\textbf{Same-Generation}
is a canonical \emph{bilinear} recursive query, expressed as:
\[
 \textit{sg}(X, Y) \leftarrow \textit{parent}(X, P_x), \textit{parent}(Y, P_y), \textit{sg}(P_x, P_y).
\]
Conceptually, this is the family-tree query. It finds pairs of nodes lying at the same generational depth, for example, siblings (one level up share a parent), cousins (two levels up share a grandparent), and so on.
Each recursive step looks up two parents in the base relation rather than one base edge, and emits the cross product of the two match sets.
The sorted-array pipeline, fused kernel, batched loop, and disjoint merge of \S\ref{sec:sorted_set}--\ref{sec:batched_loop} carry over; only two query-specific components require modification.




\textbf{Triangle Counting} is a non-recursive three-way self-join over the edge relation:
\[
  \textit{triangle}(a, b, c) \;\leftarrow\; \textit{edge}(a, b),\; \textit{edge}(b, c),\; \textit{edge}(c, a).
\]
The rule has no recursion, so the fixpoint loop reduces to a single pass. However, triangle counting stresses exactly the join bottleneck that motivated the sorted-array design: every input edge probes the edge relation by source vertex, and hub vertices serialize hash-based probes (Section~\ref{sec:data_skew}). Because triangle counting isolates the join from recursive scheduling, it provides the cleanest comparison between the sorted-array and hash-based approaches. Section~\ref{sec:eval_ablation_sorted} reports head-to-head results against mnmgJOIN's streaming hash and GPULog's materialized hash on five SNAP graphs.



%% file: ipdps/text/evaluation.tex

\section{Evaluation}
\label{sec:eval}


\textbf{Evaluation platform.}
We use an NVIDIA GeForce RTX 3060 Laptop GPU with 6 GB of memory. We intentionally choose a consumer-grade laptop GPU because \name{}'s target deployment is browser-hosted analytics on personal hardware, not datacenter accelerators.
The same machine runs both Chromium with WebGPU's Dawn/D3D12 backend (subgroup and timestamp-query support enabled) and native CUDA 12.4 with the Thrust and CUB libraries.

\textbf{Datasets.}
We use two recursive query workloads, Transitive Closure (TC) and Same-Generation (SG), on graphs from SNAP~\cite{snap} and SuiteSparse~\cite{suitesparse}.
The TC suite has 12 graphs spanning closure size ($146$K--$80$M tuples), iteration count ($11$--$426$), and degree skew (\texttt{ego-Facebook}'s 1043-degree hub).
The SG suite has 4 graphs covering the same three axes for the bilinear recursive shape ($54$--$77$ iterations, $23$K--$66$M tuples).
The triangle microbenchmark in \S\ref{sec:eval_ablation_sorted} uses 5 SNAP graphs with maximum out-degree from $343$ to $2{,}540$.

\textbf{Baselines.}
For GPU comparison we use mnmgJOIN~\cite{shovon2023} and GPULog~\cite{gpulog}, both CUDA hash-based engines.
For CPU comparison we use Soufflé as the compiled baseline and sql.js~\cite{sqljs}, Ascent WASM~\cite{ascent}, and DuckDB-WASM~\cite{duckdbwasm} as browser-bound baselines.

\textbf{Methodology.}
Production runs use $10$ warmup and $100$ timed executions per dataset, with \texttt{BATCH=30} (adaptive shrink, adaptive radix) for TC and \texttt{BATCH=10} for SG.
Larger ablation sweeps use $30$ timed executions unless stated otherwise.
End-to-end time is wall-clock query time; per-stage GPU time is measured via WebGPU timestamp queries.

\subsection{End-to-End vs.\ GPU SOTA}
\label{sec:eval_sota}

We compare \name{} against the two CUDA GPU baselines on the 12 TC graphs and the 4 SG graphs.
Tables~\ref{tab:tc_e2e} and~\ref{tab:sg_e2e} show end-to-end time and per-dataset speedup.
\name{} runs in $6.61$~s on TC and $2.18$~s on SG, compared with $15.77$~s and $3.23$~s for mnmgJOIN, and $20.14$~s and $10.21$~s for GPULog.
Cumulatively, \name{} is $2.38\times$ and $1.48\times$ faster than mnmgJOIN, and $3.05\times$ and $4.68\times$ faster than GPULog on TC and SG respectively.

\begin{table}[t]
\centering
\small
\caption{SG end-to-end time (ms) on 4 graphs.}
\label{tab:sg_e2e}
\resizebox{\columnwidth}{!}{
\begin{tabular}{l r r r r r r}
\hline
Dataset & Iter. & $|\mathit{SG}|$ & \name{} & mnmgJOIN & GPULog & vs.\ mnmgJOIN / GPULog \\
\hline
\texttt{OL.cedge}     & 56 &    285,431 &   62.317 &  101.58 &  207.37 & 1.63$\times$ / 3.32$\times$ \\
\texttt{cal.cedge}    & 58 &     23,519 &   58.142 &   52.91 &  153.49 & 0.91$\times$ / 2.64$\times$ \\
\texttt{TG.cedge}     & 54 &    608,090 &   63.325 &  108.29 &  228.60 & 1.71$\times$ / 3.61$\times$ \\
\texttt{fe\_ocean}    & 77 & 65,941,441 & 1998.453 & 2977.80 & 9614.56 & 1.49$\times$ / 4.81$\times$ \\
\hline
\end{tabular}}
\end{table}

We have three observations.
\textit{First}, performance gains scale with closure size and iteration count.
The four largest TC closures (\texttt{p2p-Gnutella09} through \texttt{SF.cedge}, $21$--$80$M tuples) account for ${\sim}78\%$ of cumulative time and yield $1.31$--$2.91\times$ wins over mnmgJOIN and $2.92$--$4.94\times$ over GPULog.
These workloads stress the two bottlenecks \name{} eliminates: hash-table contention scales with closure size and host-driven synchronization scales with iteration count.

\textit{Second}, hub-skewed graphs and long-fixpoint graphs win through different mechanisms.
\texttt{ego-Facebook}'s 1043-degree hub gives \name{} a $3.66\times$ win over GPULog by replacing hash probing with sorted range lookup.
\texttt{luxembourg\_osm}'s $426$ iterations over a $5$M closure give $2.02\times$ over mnmgJOIN despite small per-iteration work, because the host loop is amortized over batches rather than paid per stage. \textit{Third}, graphs with limited per-iteration work yield smaller gains.
\texttt{OL.cedge} ($146$K closure) loses $0.80\times$ to mnmgJOIN and \texttt{wing} ($11$ iterations) loses $0.40\times$ because their useful GPU work is too small to amortize WebGPU's fixed browser overhead.
Both losses are constant overheads rather than algorithmic regressions; they are absorbed by the $9.16$~s cumulative gap on the larger TC queries.

\textbf{SG generalization.}
The SG cumulative win over mnmgJOIN drops to $1.48\times$ (from TC's $2.38\times$) because SG's bilinear cross-product fan-out gives more compute per iteration, amortizing mnmgJOIN's hash-join overhead.
Against GPULog the gap widens to $4.68\times$ because GPULog's materialized intermediate relations grow super-linearly with the fan-out.
The single SG loss is \texttt{cal.cedge}, whose $23{,}519$-tuple closure mirrors TC's small-graph loss pattern.

\subsection{Comparison to Browser-Based Recursion Engines}
\label{sec:eval_cpu}

Browser-based deployment of recursive queries currently relies on WebAssembly CPU runtimes: database engines compiled to WASM and executed by the browser's JavaScript engine.
We compare \name{} against the three most widely-used such runtimes: sql.js~\cite{sqljs} (a SQLite WASM port), Ascent-WASM~\cite{ascent} (a Rust Datalog port), and DuckDB-WASM~\cite{duckdbwasm} (an analytical-database WASM port). All three share \name{}'s zero-install deployment context but lack GPU acceleration.
We additionally include Souffl\'e as a non-browser CPU ceiling (compiled, multi-threaded native Datalog) to bound how much of the speedup attributes to GPU acceleration versus to leaving the browser-CPU sandbox.
Table~\ref{tab:cpu_baselines} shows cumulative end-to-end time and the corresponding \name{} speedup.

\begin{table}[t]
\vspace{-10pt}
\centering
\small
\caption{Cumulative E2E time (s) for browser WASM baselines and a Souffl\'e CPU ceiling. \name{} reference: TC $6.61$, SG $2.18$.}
\label{tab:cpu_baselines}
\resizebox{\columnwidth}{!}{
\begin{tabular}{l r r r r}
\hline
Baseline & TC time & TC speedup & SG time & SG speedup \\
\hline
sql.js (browser)         & 873.00 & 132.0$\times$ & 673.64 & 308.7$\times$ \\
Ascent WASM (browser)    & 140.93 &  21.31$\times$ &  60.96 &  27.94$\times$ \\
DuckDB-WASM (browser)    &  94.92 &  14.35$\times$ &  52.97 &  24.28$\times$ \\
\hline
Souffl\'e (native CPU)   & 167.39 &  25.32$\times$ & 109.06 &  50.01$\times$ \\
\hline
\end{tabular}}
\end{table}

We have two observations.
\textit{First}, browser-based recursive query evaluation has been CPU-only before \name{}, and the gap to every WASM runtime spans more than an order of magnitude: $14$--$132\times$ on TC and $24$--$308\times$ on SG.
DuckDB-WASM and Ascent WASM are the most competitive browser baselines (both use specialised join engines), but neither can match a GPU at recursive scale.
sql.js, the most widely-deployed browser SQL engine, lags furthest because its general-purpose recursive-CTE evaluation was not designed for the iteration counts and tuple volumes of TC/SG.

\textit{Second}, even Souffl\'e (compiled, multi-threaded, and the production native CPU Datalog standard) trails \name{} by $25.32\times$ on TC and $50.01\times$ on SG, showing that the speedup is not merely a consequence of escaping the browser-CPU sandbox.
The widening from TC to SG reflects SG's heavier per-iteration work, which exposes more parallelism for the GPU to exploit relative to a multi-core CPU.

\subsection{Sorted-Array Pipeline}
\label{sec:eval_ablation_sorted}

We isolate Contribution~1 in two complementary experiments.
The within-pipeline ablation (Table~\ref{tab:ablation_sorted}) replaces sorted-array components with hash equivalents inside \name{}'s recursive engine.
The triangle microbenchmark (Table~\ref{tab:triangle}) compares the join algorithm head-to-head on a non-recursive query, removing fixpoint loop noise so the join cost is measured in isolation.

\begin{table}[t]
\vspace{-10pt}
\centering
\small
\caption{Sorted-array pipeline ablations within \name{}'s TC engine. Times in ms.}
\label{tab:ablation_sorted}
\resizebox{\columnwidth}{!}{
\begin{tabular}{l l c c c}
\hline
Design choice & Scope & Before & After & Improvement \\
\hline
Sorted-array vs.\ hash  & 6 TC graphs & 2153.45 & 876.94 & 2.46$\times$ \\
Binary-search join      & \texttt{ego-Facebook} & 42.53 & 9.69 & 4.40$\times$ \\
\hline
\end{tabular}}
\end{table}

\begin{table}[t]
\centering
\scriptsize
\caption{Triangle counting (end-to-end ms) on five SNAP graphs: \name{} sorted-array binary-search vs.\ mnmgJOIN streaming hash vs.\ GPULog materialised hash. OOM = \texttt{cudaMalloc} failure on the GPULog \texttt{two\_hop} intermediate.}
\label{tab:triangle}
\resizebox{\columnwidth}{!}{
\begin{tabular}{l r r r r r}
\hline
Dataset & edges & max-deg & \name{} & mnmgJOIN & GPULog \\
\hline
\texttt{email-Enron}      &    367,662 &  1,383 &  37.78 &    450.91 & 45,724.4 \\
\texttt{ca-AstroPh}       &    396,100 &    504 &  17.82 &    173.47 &    710.7 \\
\texttt{com-amazon}       &  1,851,744 &    549 &  26.29 &     35.20 &    605.9 \\
\texttt{soc-Slashdot0902} &    870,161 &  2,540 & 152.04 &  1,819.17 &  OOM     \\
\texttt{com-dblp}         &  2,099,732 &    343 &  38.45 &    187.83 &  1,235.3 \\
\hline
\end{tabular}}
\end{table}

We have three observations.
\textit{First}, replacing the hash path with sorted set operations gives the largest single gain inside the recursive pipeline, reducing six-dataset TC time from $2153.45$~ms to $876.94$~ms ($2.46\times$).
The improvement validates the design principle: a WebGPU engine should avoid mutable hash-table state in the recursive path.

\textit{Second}, the benefit concentrates on hub-skewed graphs.
On \texttt{ego-Facebook}, replacing the hash join with a binary-search join reduces count-and-join time from $42.53$~ms to $9.69$~ms ($4.40\times$) because the sorted range lookup avoids long hash probe chains at the 1043-degree hub.

\textit{Third}, the triangle microbenchmark confirms the gain is intrinsic to the join algorithm, not the recursive loop.
\name{} is faster than mnmgJOIN on all five graphs ($1.34$--$11.96\times$) and faster than GPULog on the four that complete ($23.0$--$1{,}210\times$).
mnmgJOIN's slowdown grows with triangle density: \texttt{ca-AstroPh} ($20.5$ triangles/edge) costs $9.74\times$ while \texttt{com-amazon} ($2.2$ triangles/edge) only $1.34\times$, because each source edge triggers a longer hash-chain walk on denser graphs.
GPULog incurs an additional materialisation cost: its \texttt{two\_hop} intermediate reaches $21$--$51$M tuples on the completed datasets, and on \texttt{soc-Slashdot0902} ($870$K edges, hub of $2{,}540$) it exceeds the $6$~GB GPU and aborts.
\name{}'s sorted-array binary-search join avoids both failure modes: per-hub work scales as $O(d \log N)$ rather than $O(d^2)$, and there is no intermediate materialisation.

\subsection{Batched Indirect-Dispatch Loop}
\label{sec:eval_ablation_batch}

\begin{figure}[t]
\centering
\vspace{-15pt}
\includegraphics[width=0.94\columnwidth]{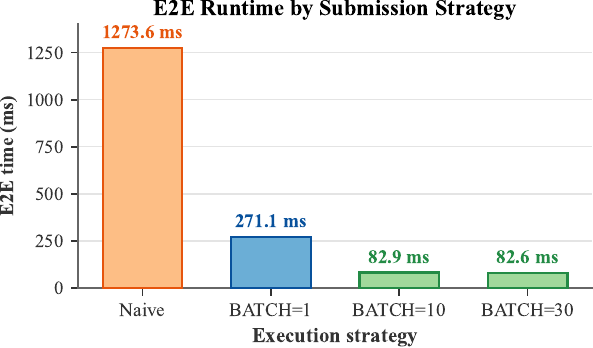}
\caption{Effect of removing host readbacks and batching WebGPU submissions on \texttt{OL.cedge}. Same kernels throughout, only submission policy varies.}
\label{fig:fence_overhead}
\vspace{-10pt}
\end{figure}

\begin{figure}[t]
\centering
\vspace{-15pt}
\includegraphics[width=0.94\columnwidth]{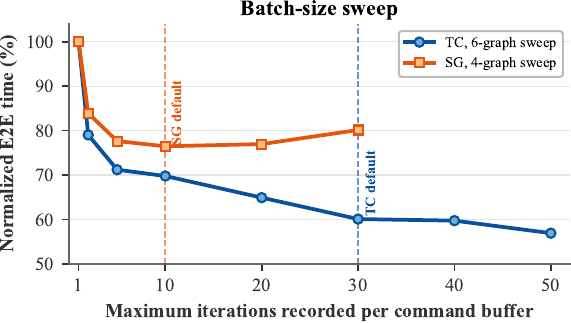}
\caption{Batch-size ablation for TC and SG. TC needs a larger adaptive batch, while SG plateaus at \texttt{BATCH=10}.}
\label{fig:batch_size_ablation}
\vspace{-10pt}
\end{figure}

We isolate Contribution~2 in two experiments.
The submission-policy ablation (Figure~\ref{fig:fence_overhead}) varies how many readbacks per iteration the host performs while keeping all kernels identical.
The batch-size sweep (Figure~\ref{fig:batch_size_ablation}) varies the number of fixpoint iterations per command submission to find the production setting.

\textbf{Batch-size selection.}
We have three observations on the submission policy.
\textit{First}, host readback is the dominant cost.
An empty WebGPU fence costs $0.176$~ms but a 4-byte readback costs $2.503$~ms, so each per-stage readback is $14.2\times$ more expensive than a fence.
A na\"ive port with three readbacks per iteration takes $1273.6$~ms on \texttt{OL.cedge}; in this mode, $93.5\%$ of end-to-end time is WebGPU synchronization rather than useful query work. \textit{Second}, indirect dispatch alone gives a $4.70\times$ speedup.
Keeping each iteration in one command buffer and reading back only the termination size (\texttt{BATCH=1}) drops time to $271.1$~ms, because indirect dispatch removes the mid-iteration readbacks used to size dependent stages. \textit{Third}, multi-iteration batching gives another $3.27\times$ on top.
\texttt{BATCH=10} drops time to $82.9$~ms and \texttt{BATCH=30} to $82.6$~ms, a cumulative $15.4\times$ reduction over the na\"ive port.
Even one readback per iteration is too expensive, so batching multiple iterations into one submission is necessary to amortize the round trip.

\textbf{Batch-size selection.}
For TC, no single fixed batch size is best across all graphs.
Short hub-heavy graphs prefer \texttt{BATCH=10--30}, medium graphs prefer \texttt{BATCH=30--40}, and long fixpoints keep improving up to \texttt{BATCH=50}.
We select \texttt{BATCH=30} as the production default because it is the largest tested setting that improves all six representative TC graphs over \texttt{BATCH=20}. \texttt{BATCH=40} regresses three graphs and \texttt{BATCH=50} regresses four graphs.
Adaptive shrinking adds another $2.2\%$ over fixed \texttt{BATCH=30} on the six-graph ablation, and improves the 12-dataset cumulative time by $7.8\%$ over the previous \texttt{BATCH=10} adaptive default.
For SG, \texttt{BATCH=10} is best on three of four graphs because the evaluated fixpoints are only $54$--$77$ iterations; \texttt{BATCH=30} regresses by $4.8\%$.

\subsection{Engineering Choices}
\label{sec:eval_ablation_eng}

The remaining design decisions are engineering optimizations that complement the two algorithmic contributions.
Table~\ref{tab:ablation_eng} reports each.

\begin{table}[t]
\centering
\scriptsize
\caption{Engineering optimizations within \name{}. Times in ms.}
\label{tab:ablation_eng}
\resizebox{\columnwidth}{!}{
\begin{tabular}{l l c c c}
\hline
Choice & Scope & Before & After & Improvement \\
\hline
Fused dedup/diff       & 6 TC graphs & 699.17 & 661.31 & 1.06$\times$ \\
Adaptive radix (sort)  & 6 TC graphs & 1353.75 & 860.07 & 1.57$\times$ \\
Adaptive radix (end-to-end)   & 6 TC graphs & 4180.33 & 3686.12 & 1.13$\times$ \\
Ping-pong vs.\ copy-back & 64M-element filter & 390.10 & 221.78 & 1.76$\times$ \\
\hline
\end{tabular}}
\end{table}

\textbf{Fused dedup/diff.}
Combining deduplication and set-difference into a single kernel saves $5.4\%$ on the six-dataset TC ablation by removing one kernel launch and one global-memory round-trip per iteration.

\textbf{Adaptive radix.}
Auto-selecting 16/24/32-bit radix sort by maximum vertex ID reduces sort time by $1.57\times$ but the end-to-end gain is only $1.13\times$ because sort is not the dominant stage on every dataset.
$16$-bit graphs save two of four passes ($1.81$--$2.01\times$ sort speedup), while $24$-bit graphs save one pass ($1.29$--$1.30\times$).

\textbf{Ping-pong vs.\ copy-back.}
WebGPU rejects a buffer range bound as both read-only input and writable output in the same compute pass, forcing either a copy-back fallback or a ping-pong design that alternates between two buffers.
At $64$M elements over $100$ filter passes, ping-pong takes $221.78$~ms while copy-back takes $390.10$~ms (extra copy adds $75.9\%$).
\name{} chooses ping-pong because the fixpoint loop is bandwidth-sensitive and repeats the same update pattern many times; the cost is $1{,}280$~MB of additional VRAM in the production TC buffers ($20.8\%$ of the $6$~GB GPU).

%% file: ipdps/text/portability.tex

\section{Portability (and conclusion)}
\label{sec:eval_portability}

Thus far, we have demonstrated \name{} as a high-performance web-based database engine and the first GPU-accelerated system of its kind. Beyond raw performance, the browser platform offers portability. We demonstrate this in this section by running the identical \name{} source code on macOS without modification. Specifically, we loaded the same build that produced our Windows/D3D12 results in Chrome on an M5 Pro MacBook, validating cross-platform compatibility.
We compare \name{} against other popular WebAssembly engines, \emph{sql.js}~\cite{sqljs}, \emph{DuckDB-WASM}~\cite{duckdbwasm}, and \emph{Ascent}~\cite{ascent}. On the 12-graph TC suite, \name{} finishes in $4.13$~s; DuckDB-WASM takes $50.72$~s, Ascent $78.75$~s, and sql.js $463.82$~s, $12.3\times$, $19.1\times$, and $112.2\times$ slower (Table~\ref{tab:portability_tc}).
On the 3-graph SG subset, \name{} runs in $109.89$~ms vs $273.80$~ms for DuckDB-WASM and $1{,}582.99$~ms for sql.js.

\begin{table}[t]
\centering
\scriptsize
\caption{TC suite on browser-deployable engines, M5 Pro MacBook.}
\label{tab:portability_tc}
\begin{tabular}{l r r}
\hline
Engine & End-to-End (s) & vs.\ \name{} \\
\hline
sql.js                   & $463.82$  & $112.2\times$ \\
Ascent WASM            &  $78.75$  & $19.1\times$ \\
DuckDB-WASM              &  $50.72$  & $12.3\times$ \\
\hline
\end{tabular}
\end{table}